\documentclass[twocolumn,aps,prl,superscriptaddress, unsortedaddress]{revtex4}
\usepackage{graphicx}
\usepackage{epsf}
\newcommand{\etal}{{\it et al.}}

\begin{document}
	
\title{The electronic structure of Bi$_2$Sr$_2$CaCu$_2$O$_y$ in the presence of a super-current: Flux-flow, Doppler shift and quasiparticle pockets. }
\author{M. Naamneh}
\affiliation{Department of Physics, Technion, Haifa 32000, Israel}
\author{J.C. Campuzano}
\affiliation{Department of Physics, University of Illinois at Chicago, Chicago, IL 60607}
\author{A. Kanigel}
\affiliation{Department of Physics, Technion, Haifa 32000, Israel}

\begin{abstract}
There are several  ways to turn a superconductor into a normal conductor:  increase the temperature, apply a high magnetic field, or run a large current. High-T$_c$ cuprate superconductors are unusual in the sense that experiments suggest that destroying superconductivity by heating the sample to temperatures above T$_c$ or by applying a high magnetic field result in different 'normal' states. Spectroscopic probes show that above T$_c$, in the pseudogap regime, the Fermi surface is partly gapped and there are no well-defined quasiparticles. Transport measurements, on the contrary, reveal quantum oscillations in high magnetic fields and at low temperatures, suggesting a more usual Fermi liquid state. Studying the electronic structure while suppressing superconductivity by using current, will hopefully shed new light on this problem. In type II superconductors, such as the cuprates, the resistive state created by the current is a result of vortex motion. The large dissipation in the flux-flow regime makes spectroscopic measurements in the presence of current challenging. 
We performed angle-resolved photoemission experiments in thin films of Bi$_2$Sr$_2$CaCu$_2$O$_y$ while running high-density current through the samples.  Clear evidence was found for non-uniform flux flow, leaving most of the sample volume free of mobile vortices and dissipation. The super-current changes the electronic spectrum, creating quasiparticle and quasihole pockets. The size of these pockets as a function of the current is found to be doping dependent; it depends both on the superfluid stiffness and on the strength of interactions. 
\end{abstract}

\date{\today}
\maketitle

The normal state of  high-temperature cuprate superconductors (HTSC) is very unusual. In a simple metallic superconductor (SC), superconductivity can be destroyed by either increasing the temperature or applying a high magnetic field; in both cases the result will be the same metallic normal state. For HTSCs, however, warming above T$_c$ results in a pseudo gap (PG) state with no well defined-quasiparticles (QP) and with a partially gapped Fermi-surface \cite{PRL08}  while applying a large magnetic field seems to result in a metallic state that can be described using the Fermi liquid theory \cite{Taillfer07}. 

There is a third way to destroy a SC; and that is by driving a high enough current through the SC.  
The ability to carry electrical current without dissipation is the hallmark of superconductivity, but at current densities above some critical value, the system becomes dissipative. In type II superconductors, such as the cuprates, there are at least two different critical current densities: the lower critical current density is associated with dissipation due to magnetic flux motion,  while at a much higher current density one expects the Cooper pairs to break due to the current and the sample to become normal.

In the case of the cuprates, due to their low superfluid density, there may be a third kind of critical current that is set by the current density at which the superfluid stiffness vanishes \cite{Goren_Altman,Ioffe_Millis}. The result is a state with finite pairing but with no phase stiffness.
 A similar state was suggested to describe the behaviour in the PG state above T$_c$, where thermal fluctuation may eliminate the superfluid stiffness without destroying the pairing \cite{EmeryandKivelson, Berg_Altman,Khodas_Tsvelik}. It will be very interesting to compare the electronic spectrum in these two states. A study describing an attempt to destroy phase coherence using current was recently published \cite{Adams_paper}.

\begin{figure} 
\begin{center}
\includegraphics[width=9cm]{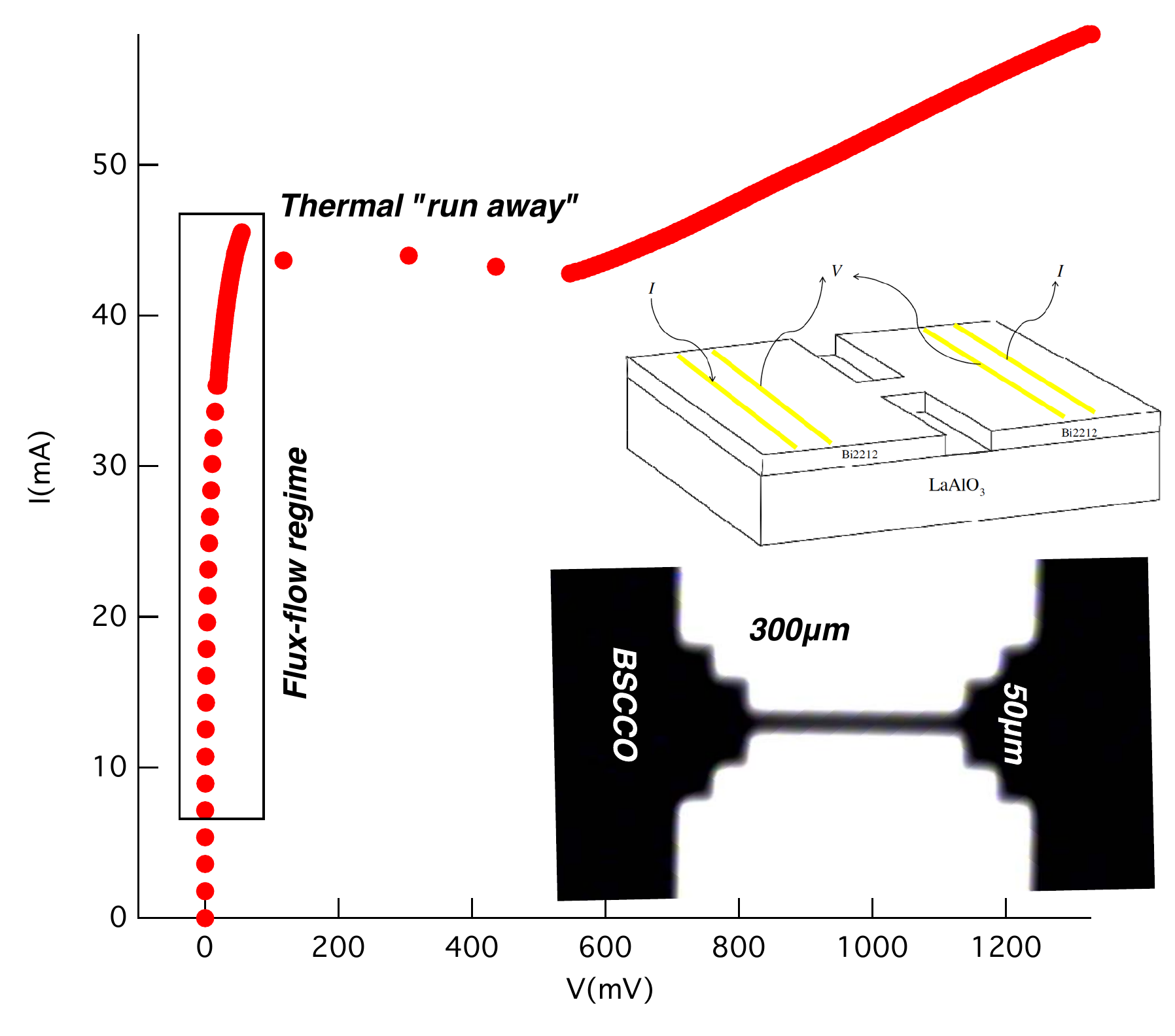}
\end{center}
\caption{ Characteristic IV of the narrow bridge used for the ARPES measurements. Three regimes are evident in the IV curve (1)  At low currents below the de-pining current no voltage is developed on the bridge. (2) Flux flows across the bridge and voltage is created (3) Above a certain current the system becomes unstable and the temperature increases rapidly due to the power dissipation. Inset: photo of the bridge and a schematic drawing of the device and the configuration of contacts.}
\label{Fig1}
\end{figure}

In this letter we present a quantitive estimate of the effect of a super-current on the spectrum of Bi$_2$Sr$_2$CaCu$_2$O$_y$ (Bi2212)  by performing ARPES measurements in the presence of electrical current in the flux-flow regime. 
To have good control over the dissipation in the sample during the experiment while running high current densities, we needed to design a suitable sample. 
We chose to work with Bi2212 thin films grown on LaAlO$_3$ substrates, and so for the purpose of this experiment, we prepared 200nm-thick Bi2212 films using DC sputtering. Using deep UV lithography, we patterned the films into narrow bridges, 50 $\mu$m in width and 300$\mu$m in length (shown in the inset of Fig. \ref{Fig1}). 
We evaporated four Au contacts onto the film; the outer contacts served as the current contacts while the inner contacts enabled  accurate measurement of the voltage drop across the bridge as the current was increased (shown also in the inset of Fig. \ref{Fig1}). The four-contact configuration is essential for several reasons: ({\it i}) it enables exact measurement of the critical current, ({\it ii}) it enables exact measurement of the power that is dissipated in the bridge and, maybe most important, ({\it iii}) it provides a simple way to measure the sample temperature and assure it remains superconducting.

Fig.\ref{Fig1} presents a typical current-voltage (IV) curve for one of our bridges. At very low currents, no voltage develops across the bridge. Once the current is increased beyond the de-pining current, magnetic flux flows and there is voltage across the bridge. All the ARPES data  were collected in the flux-flow regime. Further increasing the current  eventually leads to instability and to a thermal run-away. A systematic measurement of the doping and temperature dependence of the critical current for the same Bi2212 bridges was published recently \cite{Muntaser_PRB}.

The samples were cleaved in the ARPES UHV-chamber at a pressure lower than 1 $\times$ 10$^{-10}$ torr. To make sure that electrons are emitted only from the bridge area, where the current density is high, we used a very fine needle that was glued onto the bridge. Using these fine needles, we were able to cleave only the bridge without exposing any parts of the electrodes. 
We used a UV-laser as our photon source.  Photon energy was set to 6.52eV and the laser was focused to a $\sim$100$\mu$m diameter spot. The advantage of using low-energy photons is the improved momentum resolution obtained; the drawback is the limitation in momentum range that can be mapped. Our setup enabled the measurement of the nodal region only. Each sample was measured at a temperature at which the de-pinning current was no more than 2mA. 

\begin{figure} 
\begin{center}
\includegraphics[width=9cm]{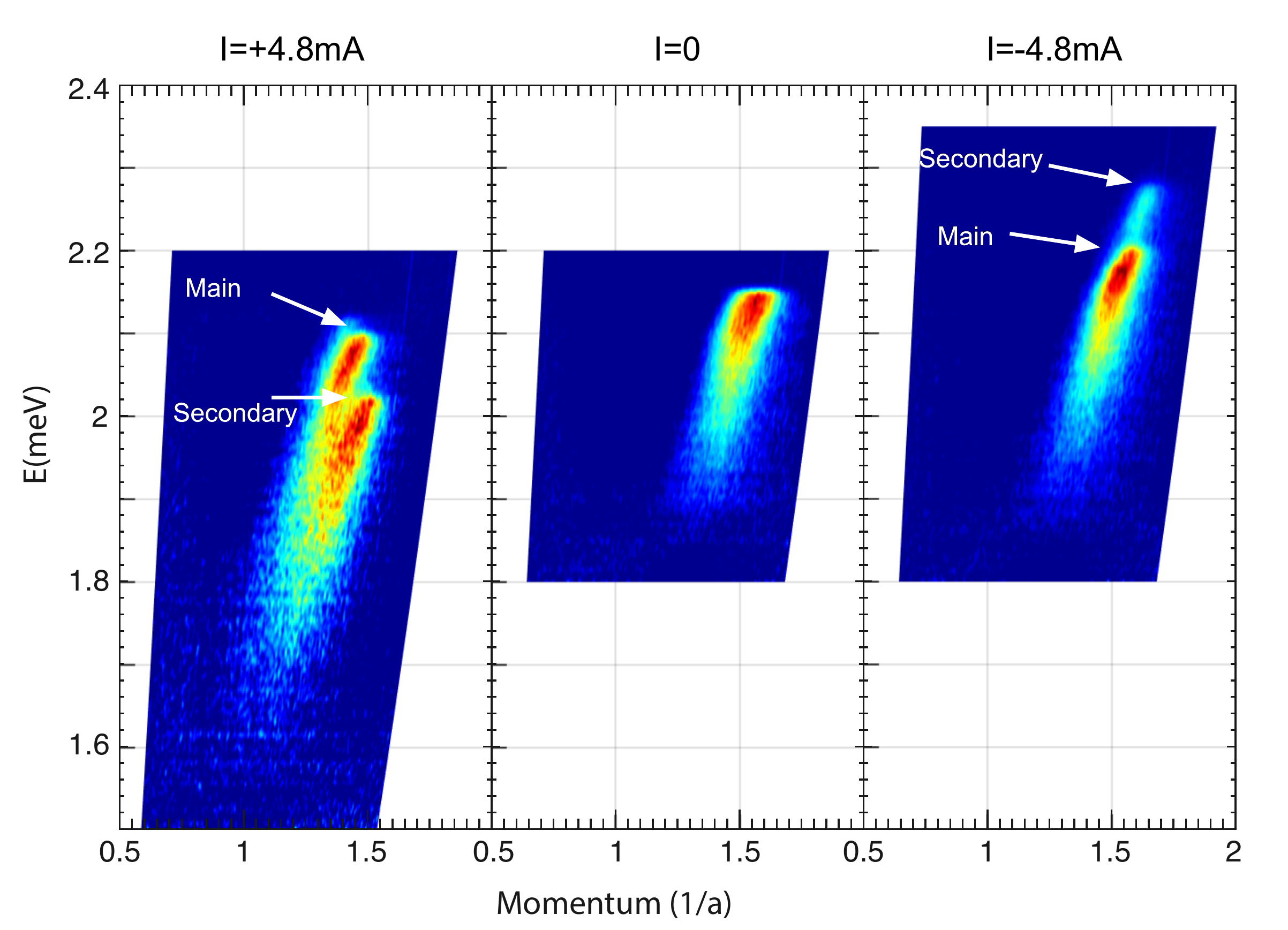}
\end{center}
\caption{ARPES spectra in the presence of current. Spectra with no current and with $\pm$4.8mA of current. In the presence of current, the spectrum is split into two distinct dispersions, which are shifted in energy. The dispersion that is shifted less with respect to the no-current dispersion is labeled "main", and the other dispersion is labeled "secondary".}
\label{Fig2}
\end{figure}

Fig.\ref{Fig2} presents the ARPES spectra of a bridge for three cases: No current, positive current of 4.8mA and negative current of 4.8mA. Very unusual behaviour is displayed: instead of the expected energy broadening, the spectra show an energy split, with no notable energy broadening.   Although the definition of the Fermi level for each dispersion is unclear, a sharp edge is clearly discernable. The dispersion that is shifted less with respect to the no-current dispersion is labeled "main", and the other dispersion is labeled "secondary". When the current is reversed, the main and secondary dispersions exchange position. The energy difference between the two dispersions is about 100meV.  In addition, the main dispersion is shifted slightly, by a few meV, relative to the Fermi level in the absence of current. 

\begin{figure*} 
\begin{center}
\includegraphics[width=15cm]{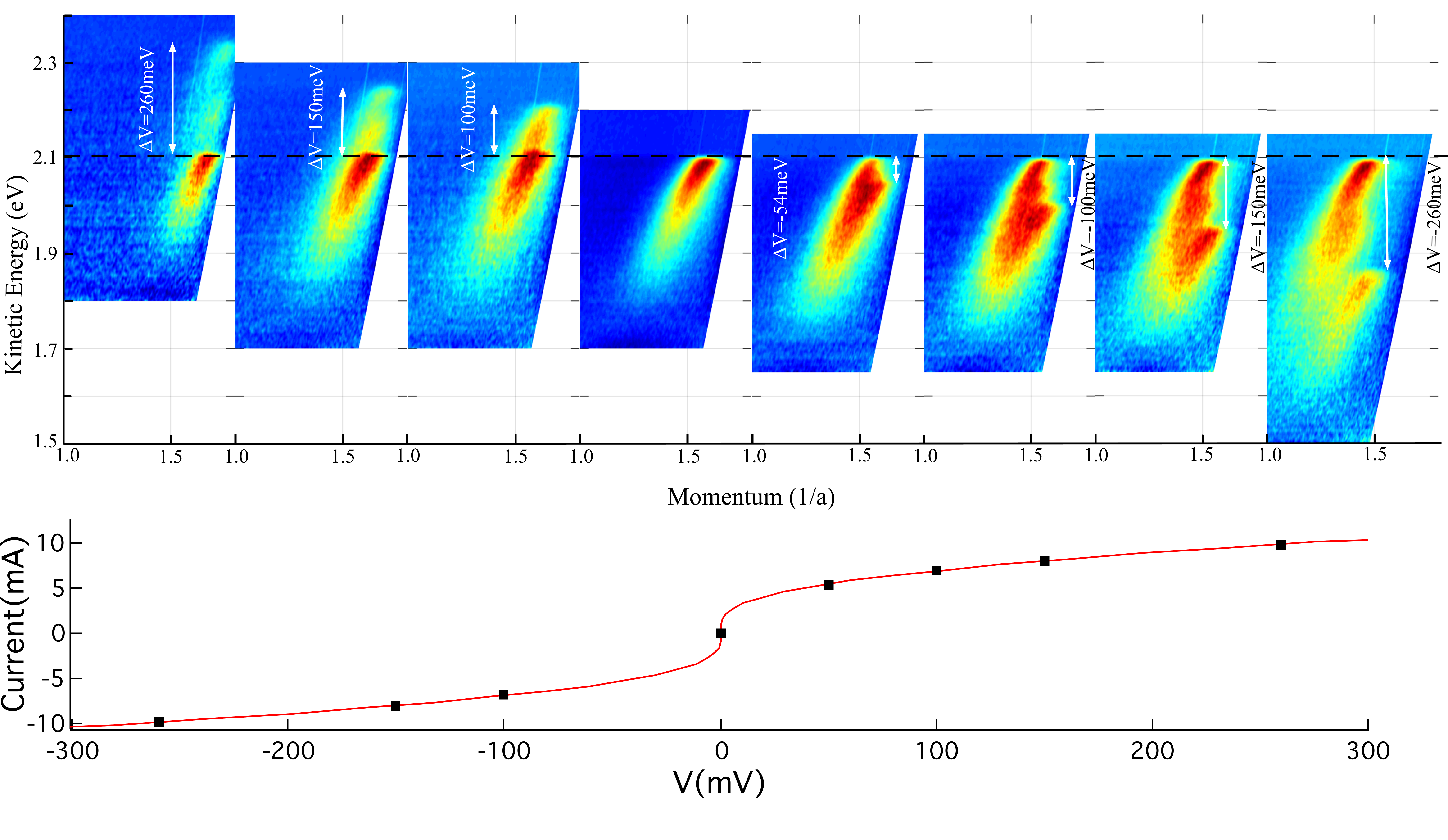}
\end{center}
\caption{Energy shift as a function of the current. The data is from a T$_c$=62K underdoped sample, measured at T=8K. Upper panel: ARPES spectra for 8 different current values. The white arrows represent the energy shift between the intensity edges of the main dispersion and the most shifted secondary dispersion. Lower panel: the black squares represent the measured energy shift as a function of the current for the 8 current values shown in the upper panel. The red line is the IV curve for the same sample measured during the ARPES measurement.}
\label{Fig3}
\end{figure*}

Next, we show data from a different sample as a function of current. The upper panel of Fig. \ref{Fig3} shows the ARPES spectra for several current densities. In this case the spectra are split into three parts. The energy difference between the different dispersions grows with the current density. The black symbols in the lower panel of Fig. \ref{Fig3} represent the energy difference between the main dispersion and the secondary dispersion for which the energy difference is maximal, as a function of the current. The red line is the IV curve measured during the ARPES measurement.

A perfect agreement is observed between the energy shift and the voltage drop across the bridge. This agreement suggests the following: for current densities larger than the de-pinning current density, vortices move through the bridge creating a voltage drop, but the flux flow is not uniform. Instead, the movement of vortices is restricted to narrow regions or channels. All of the voltage drop occurs in these channels, while most of the sample remains free of dynamic vortices and, as a result, free of phase fluctuations. In our experiment, the large majority of electrons are emitted from these parts of the sample, as the flux channels occupy a negligible volume. In this flux-flow configuration, the chemical potential does not change linearly along the current direction but instead forms sharp steps and regions of constant chemical potential. 

The chemical potential was measured at both ends of the bridge by measuring the Fermi edge of the Au contacts. The shift of the Fermi edge of the first contact with respect to the ground increases linearly with the current and is a result of contact resistance. This energy shift is identical to the energy shift of the main dispersion. The shift of the Fermi edge of the second contact equals the shift due the contact resistance plus the shift due to the voltage drop across the bridge due to the flux-flow. We can fully reproduce the IV measurements using the photoemission from the Au contacts.

All of our sample exhibited the same behavior: some of the samples displayed two dispersions, other had three. In some samples there was no energy split, but even then there was no energy broadening; in those cases there is no flux flow in the cleaved area of the bridge but rather in other parts. 
We do not believe that the splitting of the spectra is unique to our samples. Indeed, a large body of literature exists that studied the morphology of the flux flow using many different techniques in many different samples. It seems that non-uniform flux flow is the rule and not the exception \cite{Huebener_book}. 

The split in the spectra means that all phase fluctuations induced by the current are concentrated in a small part of the sample that was not probe in our ARPES experiment. These conditions create an opportunity to study SCs in the presence of very high current densities without any noticeable flux motion. Next, we describe the changes in the spectra as a result of the current.

\begin{figure} 
\begin{center}
\includegraphics[width=9cm]{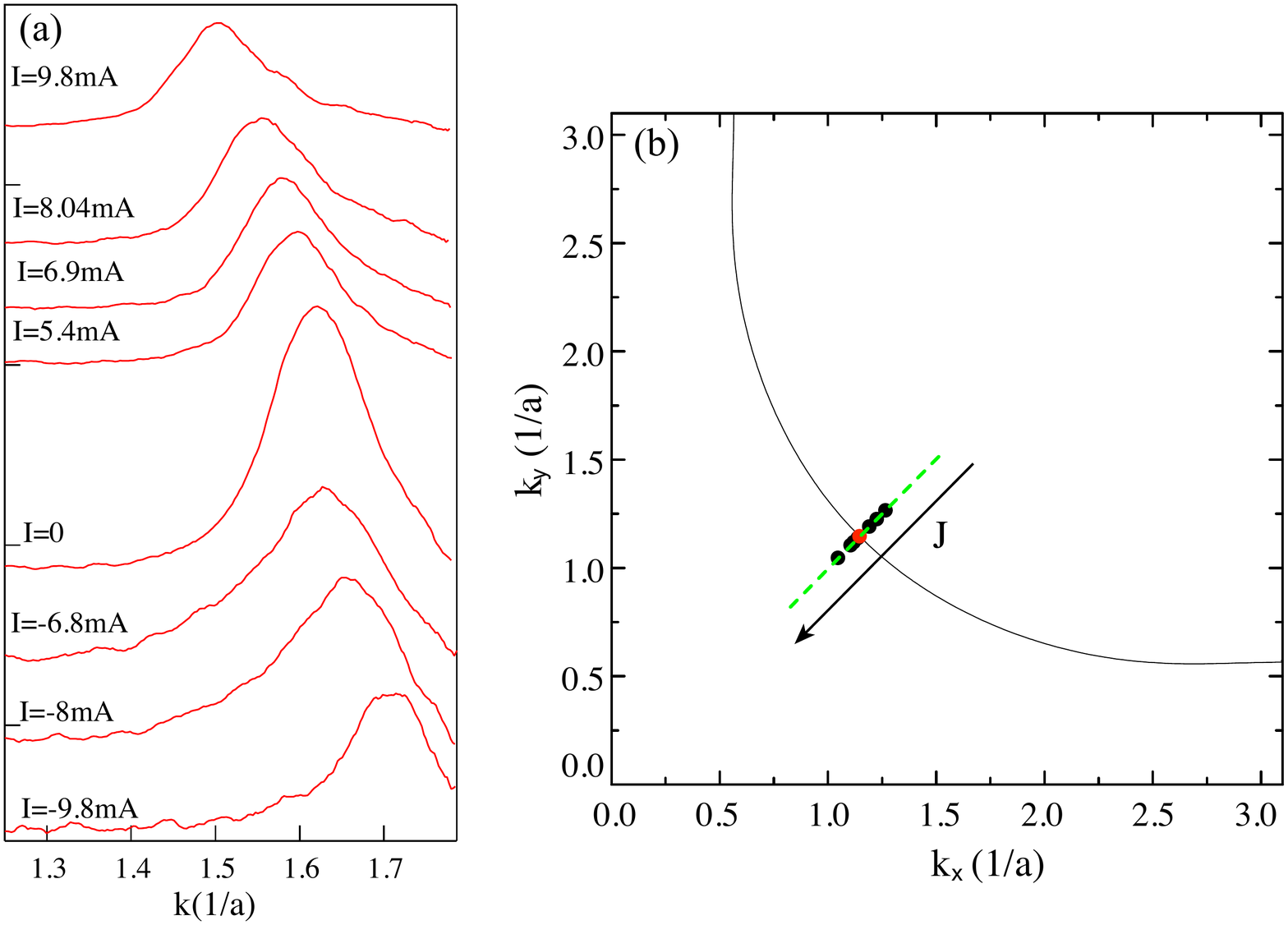}
\end{center}
\caption{Momentum shift. (a) The MDC at the intensity-edge energy for the main dispersion for several current values. The data was measured along the $\Gamma-Y$ direction. (b) Shift of the MDC peak for the same current values. The red point represents the node at zero current and the black line is the Femi-surface from a tight-binding calculation. }
\label{Fig4}
\end{figure}

Fig. \ref{Fig4}(a) presents the momentum distribution curves (MDCs) for several current values obtained at an energy that corresponds to the position of the intensity edge of the main dispersion. These data were obtained along the nodal direction. It is clear to see that the peak in MDCs shifts away from the nodal point as the current increases. 
 We also observe some broadening of the MDC peaks.  The momentum shift in the main and secondary dispersions is identical, within our experimental error.  These momentum shifts are quite significant, in Fig.\ref{Fig4}(b) we show the position of the MDC peak for different current values.  For a current of about 10mA the momentum shift is about 10\% of k$_{F}$. 

Since current is being run through the sample, thus inducing a potential difference across the sample, we expect to have stray magnetic and electric fields in the vicinity of the sample. In the supplementary material section, we estimate these stray fields and show that their effect  on the electron trajectories can account for no more than 15\% of the momentum shifts we observe. Moreover, a simple calculation shows that the magnetic field created by the current will deflect the electrons in the opposite direction to the direction found in the experiment. We therefore argue that most of the momentum shift is a result of the changes in the spectrum induced by the super-current. 

\begin{figure} 
\begin{center}
\includegraphics[width=9cm]{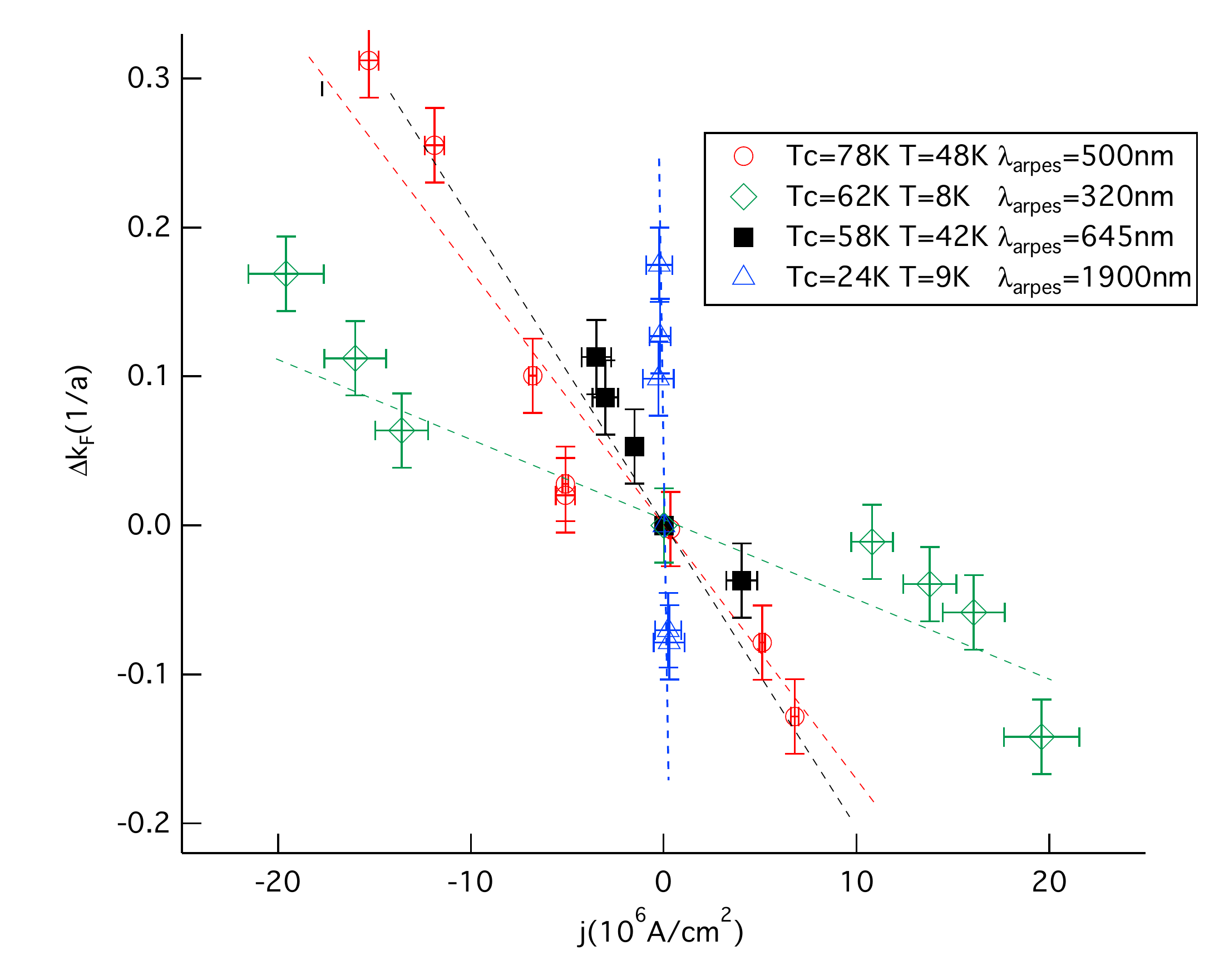}
\end{center}
\caption{Momentum shift as a function of current density for 4 different samples at different temperatures.}
\label{Fig5}
\end{figure}

Next we estimated the current density. Since the samples are cleaved, we would not use the cross-section of the bridge to calculate the current density, and so instead we used the critical current as a gauge. We measured carefully the doping and temperature dependence of the critical-current of a set of bridges for which we measured the cross section \cite{Muntaser_PRB}. Since the de-pining current density is a fundamental property of the material, we assume that the cleaved bridges have the same critical current-density, which is measured in the ARPES chamber after cleaving.  For currents larger than the de-pining current, the current flows uniformly through the narrow bridge \cite{Zeldov_current}. 
Fig. \ref{Fig5} shows the shift in k$_F$ as a function of the current density for 4 different samples.  The shift is in the direction of the current.
What is the origin of the observed momentum shift?
 
In the presence of a super-current the quasiparticle spectrum is Doppler shifted in a unique way \cite{Thinkham}. The Bogoliubov spectrum is modified:

\begin{equation}
\small
E^{\pm}(k,Q)=\frac{\xi_{k+Q/2}-\xi_{k-Q/2}}{2}\pm \sqrt{ \Big( \frac{\xi_{k+Q/2}+\xi_{k-Q/2}}{2} \Big)^2 + \Delta(k)^2 }
\label{BdG}
\end{equation}
where $\xi_k$ is the single particle dispersion. 
The Doppler shift is momentum dependent and is proportional to the super-current wave vector, $\vec{Q}$, which represents the center-of-mass momentum of the pairs.
$Q$ is proportional to  the current density. For a layered material such as Bi2212 it makes sense to use the 2D expression:
\begin{equation}
\vec{j}=\frac{e}{\hbar} \rho_s \vec{Q}
\label{Q_def}
\end{equation}
where $\vec{j}$ is the current density and $\rho_s$ is the superfluid stiffness of the SC.  

For a d-wave SC, such as Bi2212, the Doppler shift will create particle and hole pockets around the nodal points depending on the current direction \cite{Goren_Altman}. These pockets can be measured indirectly by measuring the magnetic-field dependence of the specific-heat of a d-wave SC. The magnetic field penetrates the SC in the form of a vortex lattice. The super-current circulating the vortex cores shifts the quasiparticles, creating pockets that contribute to the specific heat, a term that is proportional to $\sqrt{H}$ \cite{Volovik}. This dependence, known as the Volovik effect, was measured in the cuprates \cite{Moler}.

For small values of $Q$ along the nodal direction, we can rewrite Eq. \ref{BdG} as: 
\begin{equation}
E^{\pm}(k,Q)=v_Fk \pm \frac{1}{2} v_F Q
\label{nodal_shift}
\end{equation}
where $v_F$ is the Fermi velocity along the nodal direction. The size of the pocket created by the current along the nodal direction is Q.  Eqs. \ref{Q_def} and \ref{nodal_shift} show that for a given current density, the size of the pocket is inversely proportional to the superfluid density, which depends on both the doping and the temperature.

It is tempting to think of the momentum shift we find as a direct measurement of the pocket size. The slope of a linear fit to the data in Fig. \ref{Fig5}  is given by $\frac{\hbar d}{2e \rho_0}$ where $d$ is the c-axis lattice parameter of Bi2212. The penetration depth, $\lambda$, can now be calculated using the expression for the 2D superfluid stiffness: $$\rho_0=\frac{\hbar^2 d}{4 \mu_0 e^2 \lambda^2}$$ The values we obtained are shown in the legend of Fig.\ref{Fig5}. These values are of the same order of magnitude as values found in measurements of the absolute value of the penetration depth of Bi2212 using different techniques \cite{lambda1,lambda2,lambda3}.

Since the ARPES intensity is proportional to the spectral function, a better understating of the data requires that we consider the effect of the super-current on the spectral function. It was shown \cite{FranzMillis} that for small current densities and keeping only terms that are linear in Q, the spectral function in the presence of current can be written as $A_Q (\vec{k} , \omega) = A_0 (\vec{k}-\vec{Q}, \omega- \frac{\hbar^2}{2m} \vec{k} \cdot \vec{Q} )$, where $A_0$ is the spectral function without current. We see that in addition to the Doppler shift there is also a momentum boost, which completely cancels out the Doppler shift along the nodal direction, and so we should not expect a momentum shift at all along that direction \cite{Berg_Altman}.   
 
A possible explanation for the observed shift are the interactions, which should not be neglected. Interactions can re-normalize the current carried by the quasiparticles by a factor $\alpha$ \cite{Millis_Larkin, WenLee80}, and so in Eq. \ref{nodal_shift} we should have, in fact, used  the re-normalized term $ \frac{1}{2} \alpha V_F Q$ to describe the Doppler shift. If $\alpha < 1$, the Doppler shift will no longer be cancelled by the momentum boost and k$_F$ will be shifted. See the supplementary material for more details.

 The London penetration depth as a function of the temperature and the doping level of the same Bi2212 films, was measured using a mutual inductance technique \cite{jee}.  We found that the penetration depth extracted from the momentum shift was shorter than the values presented in Ref. \cite{jee}. The current re-normalization parameter can be estimated from the ratio of these values, $\alpha= 1- (\frac{\lambda_{ARPES}}{\lambda_{London} })^2$. For the T$_c$=78K sample we found $\alpha = 0.4 \pm 0.1$, for the moderately underdoped samples (T$_c$=58K and 62K) $\alpha = 0.6 \pm 0.1$, and for the extremely underdoped sample we find a very small value of $\alpha=0.1 \pm 0.1$ was obtained. The results are in reasonable agreement with theory \cite{Lee_alpha,Goren_Altman} and with experimental estimates \cite{Mesot,May}.  

Fig. \ref{Fig5} also reveals that for all of the samples shown, the dependence of the momentum shift dependence on the current density is not exactly linear; for high current densities it seems that $\frac{dk_F}{dj}$ increases, indicating that the superfluid density decreases. This is in agreement with theory \cite{Ioffe_Millis, Goren_Altman}, which predicts that $\rho_s=\rho_0-\frac{2 \alpha^3 v_F^2 }{\pi v_{\Delta}}Q$. In the present experiment we were not able to reach the critical current density at which $\rho_s$ is expected to vanish, as this would require increasing the current density by an order of magnitude.  We hope to accomplish this in the future, using smaller bridges. In addition, all of the data presented is limited to the nodal direction; we hope in the future to be able to measure entire pockets by extending the momentum range we can cover. 

We have shown that ARPES in the presence of current can be a useful way to understand the dynamics of flux in HTSCs and can be used to measure basic properties such as the super-fluid density and the current re-normalizaion parameter. We hope that our work will encourage others to measure different systems using similar methods.

This work was supported by the BSF, Grant No. 2010313.
We acknowledge fruitful discussions with Lilach Goren, Ehud Altman, Erez Berg, Netanel Lindner and Boris Shapira.

\newpage
\renewcommand\thefigure{S\arabic{figure}}
\setcounter{figure}{0}    

\section{Supplementary material}
\subsection{Stray magnetic and electric fields} 
Running current through a sample will create electromagnetic fields in the vicinity of the sample and these can deflect the photo-electrons on the way to the analyzer. A full  analysis of the electrons' trajectories in the vacuum chamber can be quite complicated and so we have estimated the contributions of the magnetic and electric fields separately. 
For the magnetic field we calculated the defalcation of an electron leaving the sample in normal emission, at the analyzer entrance slit. We calculated the field of a slab 50$\mu$m wide, 200nm thick and 2mm long carrying 10mA, which is the highest current we used in the experiments. We ignored the return current reduces the magnetic field. We found that the deflection under such conditions is 0.8Deg for an electron with 2.1eV kinetic energy. This is less than 15\% of the deflection we found in the experiment for the sample with the smallest deflection (largest superfluid stiffness).  

Estimating the effect of the electric field created by the current is more difficult. There are no electric fields within the SC; fields can only develop across the vortex channels where there are sharp changes in the chemical potential. We used Simion to calculate the electron trajectories in the case of one such channel 1$\mu$ wide. We assumed the voltage across the channel is 300meV, larger than any voltage drop in our experiments. Simion solves the electron's equations of motion using specific boundary conditions. We used the real dimensions of the sample, chamber and electron analyzer.  The maximal deflection for 2.1eV electrons was 0.6Deg, much smaller than the deflections we observed in the experiment.

\begin{figure} 
\begin{center}
\includegraphics[width=9cm]{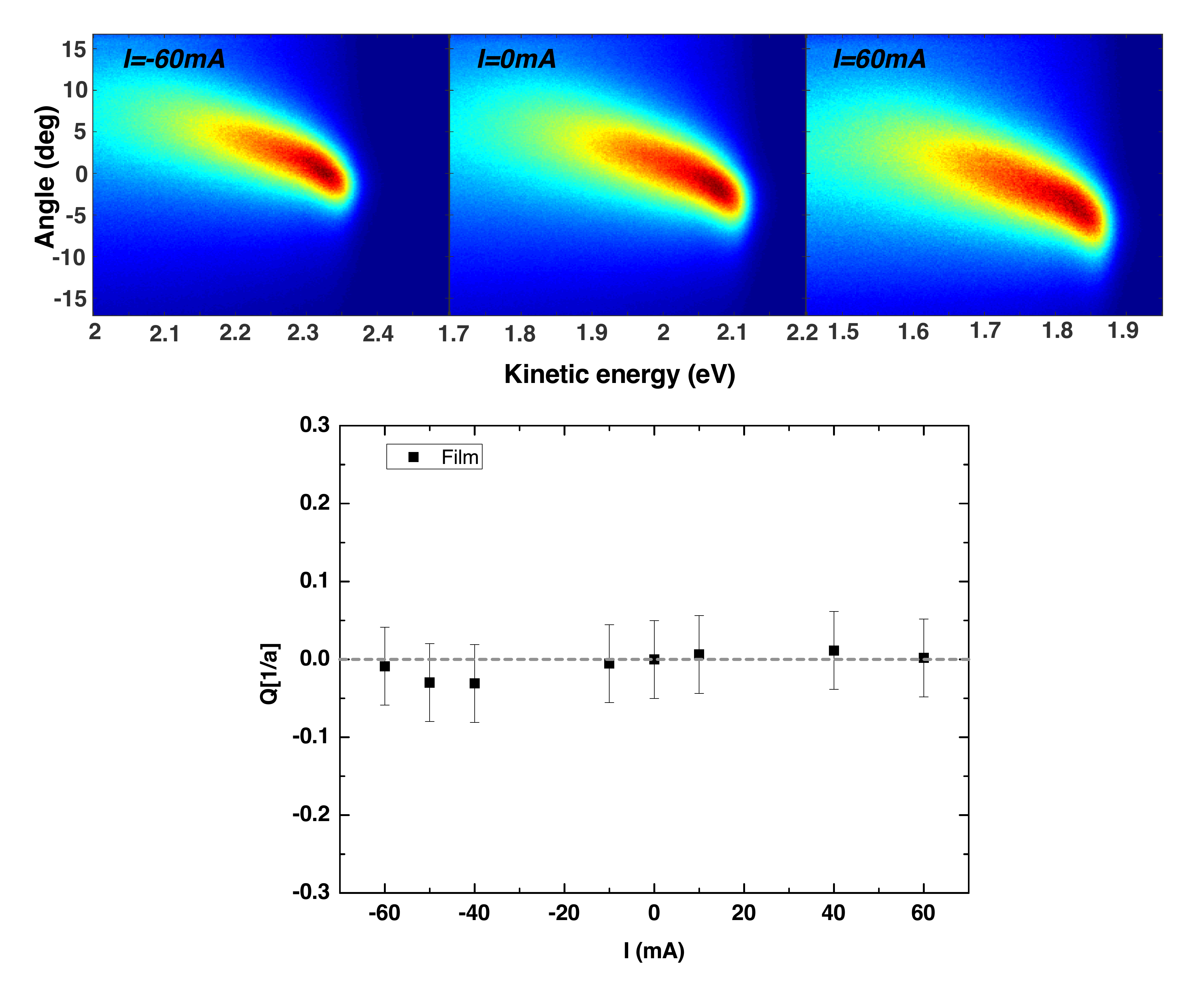}
\end{center}
\caption{Momentum shift as a function of the current for an un-patterned film.}
\label{Film}
\end{figure}

To further test the direct effect of the current on the photo-electrons we conducted the following experiment: we prepared an optimally doped film and evaporated the 4 Au contacts on it. We did not, however, patterned it into a narrow bridge. The cross section of the film was1000 larger than the cross section of the narrow bridge so we do not expect any momentum shift due to the low current density.  We ran current through the film up to 60mA. The contact resistance was such that at 60mA the voltage that developed on the current contact was about 250meV at 60mA. The results are shown in Fig. \ref{Film}. 

In the upper part of the figure we show the raw detector images for zero current and for $\pm$60mA. There is a shift in the dispersion when we plot the emission angle vs. kinetic energy.  When the angle is translated to momentum, however, no change is seen in k$_F$ of the node as a function of the current as shown in lower part of Fig. \ref{Film}. The reason for the "disappearing" of the angular shift is the large shift in the kinetic energy.

\subsection{Heating and temperature effects} 

All our measurements were done in the flux-flow regime where the samples are dissipative. The maximal power supplied during the ARPES measurements was 10mA $\times$ 500mV =  5mW, this is relatively low power  compared with the cooling power of the LHe cryostat, which is well above 1W at these temperatures. Nevertheless, heat is transferred to the cold-finger through the insulating substrate, and temperature gradients can develop. 
Since we can run a real four-terminal measurement of the IV curve of the sample during the ARPES measurements, we know the exact temperature of the bridge during the experiments. Unlike a thermometer, which measures the temperature of the sample holder, the IV curve gives an accurate measurement of the temperature of the Bi2212 film itself. We did not find any significant change in the temperature during our measurements.

\begin{figure} 
\begin{center}
\includegraphics[width=9cm]{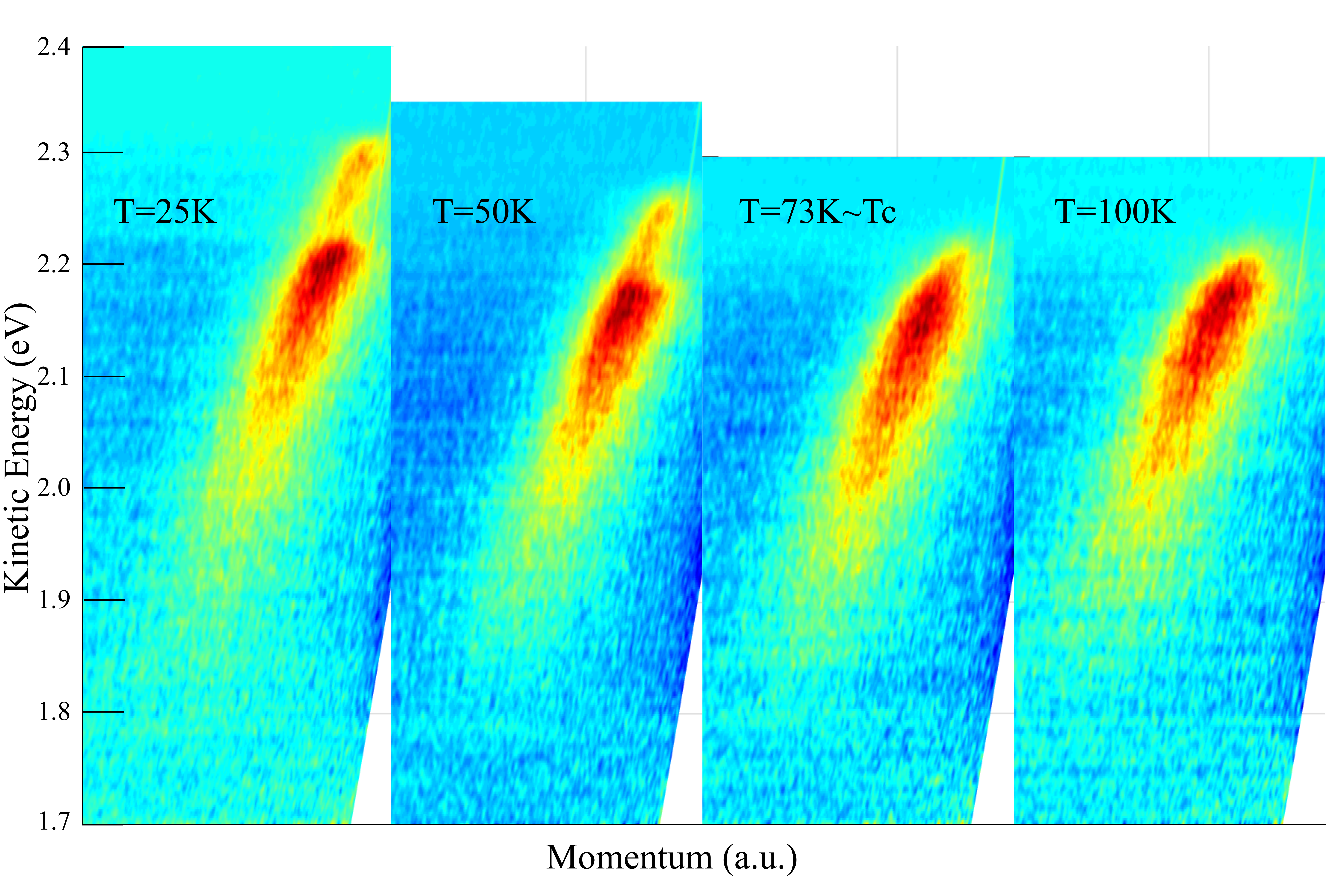}
\end{center}
\caption{Detector images for a T$_c$=73K sample as a function of the temperature.}
\label{T_scan}
\end{figure}

Fig. \ref{T_scan} presents data for a underdoped (T$_c$=73K) bridge measured at several temperatures. These measurements were done at different currents chosen such that the voltage across the bridge was 100meV. The split could be observed only below T$_c$, where even at T=73K$\sim$T$_c$ a small secondary dispersion can be observed.  Several interesting things can be learned from this image: at 25K and 50K, deep inside the SC state, the energy difference between the main and secondary dispersion is the same, 100meV, which is the entire voltage drop across the bridge. Note that the position of the main dispersion is different because at 25K since more current is needed for the bridge to develop 100meV and this leads to a larger voltage drop on the current contact. At T$_c$, the energy difference between the dispersions is much smaller although the voltage across the bridge is the same, 100meV. The reason for this is that most of the resistance at T$_c$ is not related to flux-flow but rather to normal parts of the sample. Above T$_c$, the sample is a plain resistor. The broadening due to the variation of the electro-chemical potential across the bridge is not significant. The 100meV develops across a 300$\mu$m bridge, so an emission spot of about 100$\mu$m will result in a broadening of no more than 30meV, which is consistent with our data. 

\begin{figure*} 
\begin{center}
\includegraphics[width=18cm]{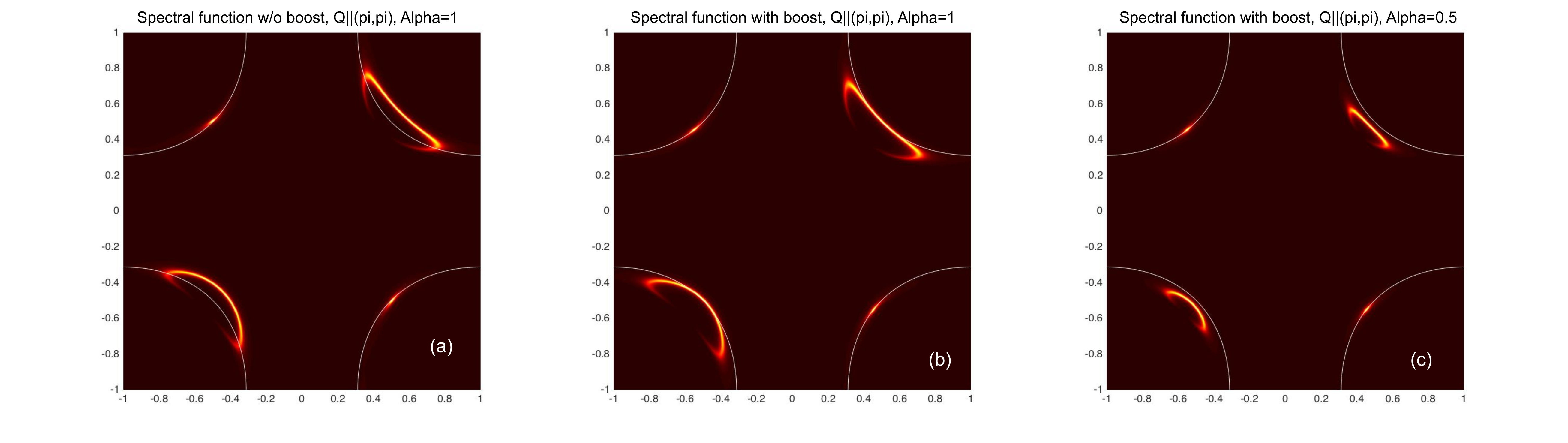}
\end{center}
\caption{The spectral function with current. (a) Calculation of the spectral function for the case of a current in the ($\pi,\pi$) direction with no momentum boost. (b) The same as (a) with the momentum boost. (c) The spectral function for the case of a re-normalized current ($\alpha = 0.5$). }
\label{Ak}
\end{figure*}

\subsection{Spectral function in the presence of current}

The ARPES intensity is proportional to the spectral function that is given by the imaginary part of Green's function. We use the usual BCS Green's function to calculate the spectral function:
$$G^{-1}(k,\omega)_0= \omega - \varepsilon_k + i\Gamma -\frac{\Delta_k^2}{\omega + \varepsilon_k}$$
where $\varepsilon_k$ is the tight-binding dispersion, $\Delta_k$ is the d-wave gap and $\Gamma$ is a single-particle scattering-rate. 

Using this approximation, the ARPES data may be fit below T$_c$ \cite{Phenomenology_PRB} as was shown before.  As shown by Franz and Millis \cite{FranzMillis} a uniform super-current will modifies the Green's function as follows: 
$$G_Q(k,\omega)=G_0(\vec{k}-\vec{Q}/2,\omega- \frac{\alpha \hbar^2}{2m} \vec{k} \cdot \vec{Q}) $$
where $\alpha$ is current re-normalization parameter. The Doppler shift in d-wave superconductors creates particle and hole pockets. In Fig.\ref{Ak}  we show the spectral function at the Fermi level for the case where
 $\vec{Q}$ is in the ($\pi,\pi$) direction. Panel (a) presents the spectral function, ignoring the momentum boost (without taking $k$ to $k-Q$) for $\alpha=1$. In this case we get pockets of size Q centred around the original nodes. The spectral weight is higher on one side of the pockets and, along the nodal direction in particular, the spectral weight vanishes on the "dark" side of the pockets. In Panel (b) we add the momentum boost, but maintain $\alpha=1$. As can be seen, the center of the pockets is shifted by Q/2 leaving k$_F$ of the "bright" side of the pockets exactly at the nodes.  Panel (c) shows the spectral function for $\alpha=0.5$; the result is smaller pockets of size $\alpha Q$. Since the shift due to the momentum boost remains $Q/2$ we do observe a shift in k$_F$. 

\end{document}